\documentclass[a4paper,oneside,11pt]{article}
\usepackage{cprform}
\usepackage{amsmath,amstext,amsfonts,amsbsy,amssymb,amscd,bbm,epsfig,lscape}

\usepackage{epsfig,macros,cite}
\usepackage{amssymb}
\usepackage{subfigure}
\usepackage{sidecap}
\usepackage{wrapfig}
\usepackage{float}
\usepackage{rotating}
\usepackage[table]{xcolor}
\usepackage{booktabs}
\usepackage{multirow}
\usepackage{hhline}
\usepackage{array}

\newcommand{\ci}[1]{\boldsymbol{#1}}

\newcommand{\Dslash}{\relax{\kern+.25em / \kern-.70em D}}

\newcommand{\sign}{\rm sign}

\newcommand{\Real}{\relax{\mathsf{\Gamma\kern-.35em R}}}
\newcommand{\Int}{\relax{\mathsf{Z\kern-.40em Z}}}

\newcommand{\obar}[1]{\kern3pt\overline{\kern-2pt #1\kern-0pt}\kern1pt}

\newcommand{\corrbar}[1]{\kern3pt\overline{\kern-2pt #1\kern-0pt}\kern1pt}

\newcommand{\oVApAVren}[1]{\kern3pt\overline{\kern-2pt #1\kern-0pt}\kern1pt_{\rm\scriptscriptstyle VA+AV;s}}

\newcommand{\Kstar}{K^*}

\newcommand{\fT}{f_{\rm\scriptscriptstyle T}}

\newcommand{\zbar}{\kern3pt\overline{\kern-2pt Z\kern-0pt}\kern1pt}

\newcommand{\zbarVApAV}[1]{\kern3pt\overline{\kern-2pt Z\kern-0pt}\kern1pt_{\rm\scriptscriptstyle VA+AV #1}}

\newcommand{\cO}{{\cal O}}

\begin{document}
\bibliographystyle{mybibstyle}


\begin{titlepage}


\vspace*{-30truemm}
\begin{flushright}
ROM2F/2011/04, \\ICCUB-11-129, UB-ECM-PF-11-48\\
\today
\end{flushright}
\vspace{5truemm}


\centerline{{\Large $\Kstar$ vector and tensor couplings from $N_f = 2$ tmQCD}}
\vskip 9 true mm
\vskip -2 true mm
\begin{center}
\epsfig{figure=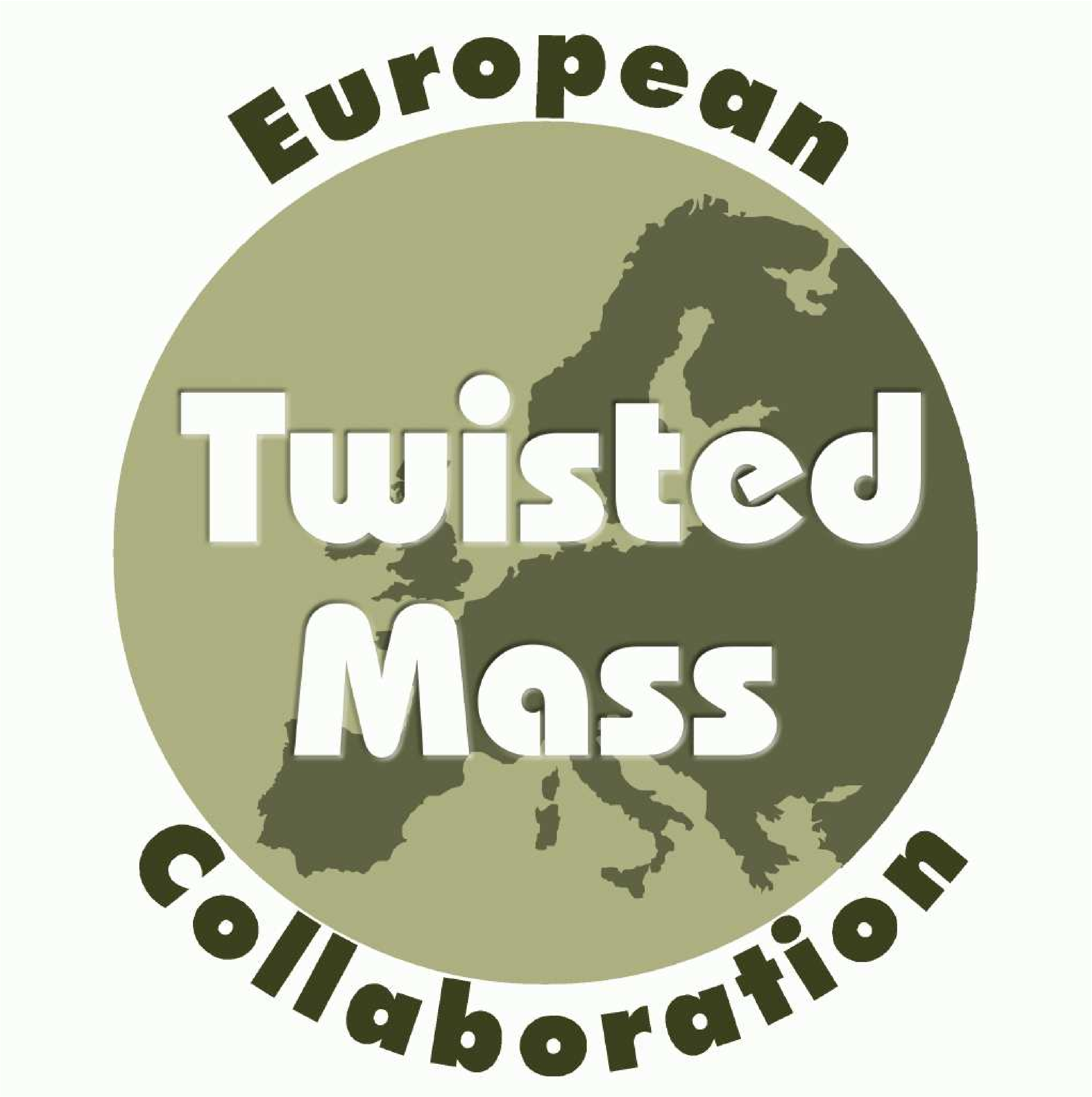, width=22 true mm}\\
\end{center}
\centerline{\Bigrm ETMC}
\vskip 4 true mm
\centerline{\bigrm  P.~Dimopoulos$^a$, 
F.~Mescia$^c$,
and A.~Vladikas$^d$}
\vskip 4 true mm
\centerline{\it $^a$ Dipartimento di Fisica, Universit\`a di Roma ``Tor Vergata''}
\centerline{\it Via della Ricerca Scientifica 1, I-00133 Rome, Italy}
\vskip 3 true mm
\centerline{\it $^c$ Departament dÕEstructura i Constituents de la Mat\'eria and Institut
de Ci\'encies del Cosmos,}
\centerline{\it Universitat de Barcelona, Diagonal 647,}
\centerline{\it E-08028 Barcelona, Spain}
\vskip 3 true mm
\centerline{\it $^d$ INFN, Sezione di ``Tor Vergata"}
\centerline{\it c/o Dipartimento di Fisica, Universit\`a di Roma ``Tor Vergata''}
\centerline{\it Via della Ricerca Scientifica 1, I-00133 Rome, Italy}
\vskip 10 true mm

\thicktablerule
\vskip 3 true mm
\noindent{\tenbf Abstract}
\vskip 1 true mm
\noindent
{\tenrm  The mass $m_{\Kstar}$ and vector coupling $f_{\Kstar}$ of the $\Kstar$-meson, as well as the ratio of the tensor to vector couplings $\dfrac{f_T}{f_V}\Big \vert_{\Kstar}$,  are computed in lattice QCD.
Our simulations are performed in a partially quenched setup, with two dynamical (sea) Wilson quark flavours, having
a maximally twisted mass term. Valence quarks are either of the standard or the Osterwalder-Seiler
maximally twisted variety. Results obtained at three values of the lattice spacing are extrapolated to the continuum, giving
$m_{\Kstar} = 981(33) {\rm MeV}$, $f_{\Kstar} = 240(18) {\rm MeV}$ and $\dfrac{f_{T}(2\,{\rm GeV})}{f_{V}}\Big \vert_{\Kstar} = 0.704(41)$.}
\vskip 3 true mm
\thicktablerule
\vspace{10truemm}
\eject
\end{titlepage}

\section{Basics}
\label{sec:intro}

The aim of the present letter is to present novel lattice results for the mass of the $\Kstar$-meson, as well as its vector and tensor couplings ($f_V$ and $f_T$ respectively), defined in Euclidean space-time as follows:
\begin{eqnarray}
\langle 0 \vert V_j \vert \Kstar;\lambda \rangle \,\, &=& \,\,  \, -i f_V \epsilon^\lambda_j \, m_{\Kstar} \,\,\, ,
\\
\langle 0 \vert T_{0j} \vert \Kstar;\lambda \rangle \,\, &=& \,\, -i \, f_T \epsilon^\lambda_j \, m_{\Kstar} \,\,\, .
\end{eqnarray}
In the above expressions, $V_j = \bar s \gamma_j d$ is the vector current (spatial components only; $j=1,2,3$), $T_{0j} = i \bar s \sigma_{0j} d$ is the tensor bilinear operator (temporal component), and $\epsilon^\lambda_j$ denotes the polarization vector.

Our results are based on simulations of the ETM Collaboration (ETMC)~\cite{Baron:2009wt}, with $N_f =2$ dynamical flavours (sea quarks) and ``lightish" pseudoscalar meson masses in the range $280 \, {\rm MeV} < m_{\rm PS} < 550 \, {\rm MeV}$. With three lattice spacings ($a = 0.065~§{\rm fm}$, $0.085~{\rm fm}$ and 0.1~fm) we are able to extrapolate our results to the continuum limit. Our simulations are performed with the tree-level Symanzik improved gauge action. For the quark fields we adopt a somewhat different regularization for sea and valence quarks. The sea quark lattice action is the so-called maximally twisted standard tmQCD (referred to as ``standard tmQCD case")~\cite{Frezzotti:2000nk}. The $N_f =2$ light sea quark flavours form a flavour doublet $\bar \chi = (\bar u~,~\bar d)$ and the fermion lattice Lagrangian in the so-called ``twisted basis" is given by 
\be
{\cal L}_{\rm tm} \,\, = \,\, \bar \chi \Big [ D_W \,\, + i \mu_q \gamma_5 \tau^3 \Big ] \chi \,\, ,
\label{eq:Ltm}
\ee
where $\tau^3$ is the isospin Pauli matrix and $D_W$ denotes the critical Wilson-Dirac operator. By ``critical" we mean that, besides the standard kinetic and Wilson terms, the operator also includes a standard, non-twisted mass term, tuned at the critical value of the quark mass ($\kappa_{\rm cr}$ in the language of the hopping parameter), so as to ensure maximal twist. With only two light dynamical flavours, strangeness clearly enters the game in a partially quenched context. For the valence quarks we use the so-called Osterwalder-Seiler variant of tmQCD, which consists in maximally twisted flavours which, unlike the standard tmQCD case, are not combined into isospin doublets:
\be
{\cal L}_{\rm OS} \,\, = \,\, \sum_{f=d,s} \bar q_f \Big [ D_W \,\, + i \mu_f \gamma_5 \Big ] q_f \,\, ,
\label{eq:LOS}
\ee
with $\sign(\mu_f) = \pm 1$ (see below for details).
This action, introduced in ref.~\cite{Osterwalder:1977pc} and implicitly used in~\cite{tmqcd:DIrule}, has been studied in detail in ref.~\cite{Frezzotti:2004wz}. For the case in hand (i.e. $\Kstar$-related quantities) we only need down- and strange-quark flavours in the valence sector. Note that the choice of maximally twisted sea and valence quarks implies $\cO(a)$-improvement of the physical quantities (i.e. the so-called automatic improvement of masses, correlation functions and matrix elements)~\cite{FrezzoRoss1}. Thus unitarity violation, which plagues any partially quenched theory at finite lattice spacing, is an $\cO(a^2)$ effect.

The sign of $\mu_s$ may be that of $\mu_d$ or its opposite. We conventionally refer to the setup in which $\sign(\mu_d) = -\sign(\mu_s) $ as the ``standard twisted mass regularization" (denoted by tm) and the setup with $\sign(\mu_d) = \sign(\mu_s) $ as the ``Osterwalder-Seiler regularization" (denoted as OS). Quenched pseudoscalar masses and decay constants in tm- and OS-setups have already been studied~\cite{Dimopoulos:2007cn,Dimopoulos:2009es}.

The continuum operators of interest are expressed, in terms of their lattice counterparts, as follows:
\begin{eqnarray}
\label{eq:Vcont}
V_\mu^{\rm cont} \,\, &=& \,\, Z_A \, A_\mu^{\rm tm} \, + \, \cO(a^2) \,\, = \,\, Z_V \, V_\mu^{\rm OS} \, + \, \cO(a^2) \,\,\, , \\
T_{\mu\nu}^{\rm cont} \,\, &=& \,\, Z_T \, T_{\mu\nu}^{\rm tm} \, + \, \cO(a^2) \,\, = \,\, Z_T \, \tilde T_{\mu\nu}^{\rm OS} \, + \, \cO(a^2) \,\,\, ,
\label{eq:Tcont}
\end{eqnarray}
where $\tilde T_{\mu\nu} = \epsilon_{\mu\nu\rho\sigma} T_{\rho\sigma}$. The vector and axial currents are normalized by the scale independent factors $Z_V$ and $Z_A$, while $Z_T \equiv Z_T(\mu)$ runs with a renormalization scale $\mu$ (i.e. it is defined in a given renormalization scheme).

The vector boson mass, $m_V$, as well as $f_V$ and $f_T$, are obtained form two-point correlation functions at zero spatial momenta and large time separations. These are defined in the continuum (Euclidean space-time) as
\begin{eqnarray}
C_V^{\rm cont}(x^0) \,\,&\equiv& \,\, \dfrac{1}{3} \sum_j \int d^3x \,\,\, \langle V_j(x) \,\,  V_j^\dagger(0) \rangle^{\rm cont} \nonumber \\
&\rightarrow& \,\,\,
\dfrac{f_V^2 m_{V}}{2} \,\, \exp[- \m_{V} T /2] \,\, \cosh\big[ m_{V} (T/2 - x^0) \big] \,\,\, ,
\label{eq:CV}
\\
C_T^{\rm cont}(x^0) \,\, &\equiv& \,\, \dfrac{1}{3} \sum_j \int d^3x \,\,\, \langle T_{0j}(x) \,\,  T_{0j}^\dagger(0) \rangle^{\rm cont} \nonumber \\
&\rightarrow& \,\,\,
\dfrac{f_T^2 m_{V}}{2} \,\, \exp[- \m_{V} T /2] \,\, \cosh\big[ m_{V} (T/2 - x^0) \big] \,\,\, .
\label{eq:CT}
\end{eqnarray}
The asymptotic expressions of the above equations correspond to the large time limit of the correlation functions (symmetrized in time), with periodic boundary conditions for the gauge fields and (anti)periodic ones for the fermion fields in the (time)space directions (i.e. $0 \ll x^0 \ll T/2$). These are actually the boundary conditions of our lattice simulations. The lattice correlation functions are related to the continuum ones as suggested by eqs.~(\ref{eq:Vcont}),(\ref{eq:Tcont}):
\begin{eqnarray}
C_V^{\rm cont}(x^0) \,\, &=& \,\, Z_A^2 \, C_A^{\rm tm}(x^0) \, + \, \cO(a^2) \,\, = \,\, Z_V^2 \, C_V^{\rm OS}(x^0) \, + \, \cO(a^2) \,\,\, , 
\label{eq:CVren} \\
C_T^{\rm cont}(x^0) \,\, &=& \,\, Z_T^2 \, C_T^{\rm tm}(x^0) \, + \, \cO(a^2) \,\, = \,\, Z_T^2 \, C_{\tilde T}^{\rm OS}(x_0) \, + \, \cO(a^2) \,\,\, .
\label{eq:CTren}
\end{eqnarray}
The meaning of the notation $C_A^{\rm tm}$, $ C_{\tilde T}^{\rm OS}$, etc. should be transparent to the reader. The ratio $f_T/f_V$ is computed from the square root of the ratio of correlations functions $C_T^{\rm cont}/C_V^{\rm cont}$, in which many systematic effects cancel. 
We compute the vector meson mass and decay constant from $C_V^{\rm cont}$ and the ratio $f_T/f_V$ from the ratio of correlation functions $C_T^{\rm cont}/C_V^{\rm cont}$. The tensor coupling $f_T$ is then obtained by multiplying $f_T/f_V$ by $f_V$.

Note that $f_V$ is a scale independent quantity, while $f_T(\mu)$ depends on the renormalization scale $\mu$, as well as the renormalization scheme. The scale and scheme dependence of the latter quantity is carried by the renormalization factor $Z_T(\mu)$; we opt for the $\msbar$-scheme and for $\mu =  2$~GeV.

\section{Results}
\label{sec:res}

ETMC has generated $N_f =  2$ configuration ensembles at four values of the inverse gauge coupling; in this work we make use of only three of them. Light mesons consist of a valence quark doublet, with twisted mass $a\mu_\ell$ equal to that of the sea quarks; $a\mu_\ell = a \mu_{\rm sea}$. Heavy-light mesons consist of a valence quark pair $(a\mu_\ell = a \mu_{\rm sea}, a\mu_h)$. As already stated, these bare quark mass parameters are chosen so as to have light pseudoscalar mesons (``pions") in the range of $280 \leq m_{\rm PS} \leq 550$~MeV and heavy-light pseudoscalar mesons (``Kaons") in the range $450 \leq m_{\rm PS} \leq 650$~MeV. The simulation parameters are gathered in Table~\ref{simuldetails}.
\begin{table}[!h]
\begin{center}
\begin{tabular}{cccclclclc}
\hline \hline
 $\beta$  &&  $a^{-4}(L^3 \times T)$ && $a\mu_{\ell}~=~a\mu_{sea}$      &&  $a\mu_{h}$ &$N_{\rm{meas}}$&   \\
\hline
3.80      &&  $24^3 \times 48$&& 0.0080, 0.0110   && 0.0165, 0.0200   & 180 &     \\
($a\sim0.1~\mbox{fm}$)          &&    &&         &&    0.0250    &&     \\
\hline
3.90      &&  $24^3 \times 48$&& 0.0040 &&   0.0150, 0.0220 &400 &    \\
          &&                  &&        &&            0.0270        &&    \\
          &&  $24^3 \times 48$&& 0.0064, 0.0085,&&   0.0150, 0.0220 & 200&    \\
          &&                  && 0.0100         &&       0.0270        &&    \\	  
3.90      &&  $32^3 \times 64$&& 0.0030, 0.0040 &&   0.0150, 0.0220 &270/170&    \\
($a\sim0.085~\mbox{fm}$) && && &&               0.0270&& \\
\hline
4.05      &&  $32^3 \times 64$&& 0.0030, 0.0060, &&   0.0120, 0.0150 &200&      \\ 
($a\sim0.065~\mbox{fm}$) &&   && 0.0080  &&           0.0180&&      \\
\hline \hline
\end{tabular}
\end{center}
\caption{Simulation details}
\label{simuldetails}
\end{table}

Our calibrations are based on earlier collaboration results. The ratio $r_0/a$, known at  each value of the gauge coupling $\beta$ from ref.~\cite{Blossier:2010cr}, allows to express our raw dimensionless data (quark masses, meson masses and decay constants) in units of $r_0$. Knowledge of the renormalization constant $Z_P$ in the $\msbar$ scheme at 2 GeV (see ref.~\cite{Constantinou:2010gr}) enables us to pass from bare quark masses to renormalized ones (again in $r_0$ units). Using only data with light valence quarks in the tm-setup, we have applied the procedure described in refs.~\cite{Baron:2009wt,Blossier:2010cr} for the determination of the physical continuum light quark mass $\mu^\msbar_{u/d}$. From the data concerning light and heavy valence quark masses in the tm-setup~\cite{Blossier:2010cr}, we determine the physical continuum strange quark mass $\mu^\msbar_s(2\, \rm{GeV})$. These quark mass values are listed in Table~\ref{Qmassvaluse}. The Sommer scale we use, based on an analysis with three values of the lattice spacing,  is $r_0 = 0.448(5)$~fm. This updates our previous $r_0$  
computation, derived with two $\beta$'s, cf. ref.~\cite{Baron:2009wt}.

We see from eqs.~(\ref{eq:CVren}) and (\ref{eq:CTren}) that we need to know the renormalization parameters $Z_V$, $Z_A$, and $Z_T$. These quantities, as well as $Z_P$, have been computed in ref.~\cite{Constantinou:2010gr}, in the RI/MOM scheme; $Z_P$ and $Z_T$ are perturbatively converted to $\msbar$. In the same work a $Z_V$ estimate, obtained from a Ward identity, is also provided. In Table~\ref{etmc_RCs} we gather the most reliable estimates of ref.~\cite{Constantinou:2010gr}, which we have used in the present analysis, as well as our estimates of the $r_0/a$ ratio.
\begin{table}[!h]
\begin{center}
\begin{tabular}{ccccccccccc}
\hline \hline
 &&  &&  &&  &&  && \\
$\beta$ && $Z_V$ &&  $Z_A$ && $Z_T^{\rm{\overline{MS}}}(2 ~\rm{GeV})$ && 
$Z_P^{\rm{\overline{MS}}}(2 ~\rm{GeV})$ && $r_0/a$ \\
 &&  &&  &&  &&  && \\
\hline
 &&  &&  &&  &&  && \\
3.80    && 0.5816(02)&&  0.746(11) && 0.733(09) && 0.411(12) && 4.54(07) \\
3.90    && 0.6103(03)&&  0.746(06) && 0.743(05) && 0.437(07) && 5.35(04) \\
4.05    && 0.6451(03)&&  0.772(06) && 0.777(06) && 0.477(06) && 6.71(04) \\
 &&  &&  &&  &&  && \\
\hline \hline
\end{tabular}
\end{center}
\caption{The renormalization parameters used in our analysis and the $r_0/a$ values at each gauge coupling. $Z_V$ is obtained from a lattice vector Ward identity, while the other renormalization constants are obtained from the RI/MOM scheme; for details see ref.~\cite{Constantinou:2010gr}.}
\label{etmc_RCs}
\end{table}

\begin{table}[!h]
\begin{center}
\begin{tabular}{cccc}
\hline \hline
&&& \\
$\mu_{u/d}^{\overline{\rm{MS}}}( 2~ \rm{GeV})$  &&& $\mu_s^{\overline{\rm{MS}}}(2~ \rm{GeV})$ \\
&&& \\
\hline
&&& \\
 3.6(2) MeV   &&& 95(6) MeV\\ 
&&& \\
\hline \hline
\end{tabular}
\end{center}
\caption{The quark mass values (in the $\msbar$ scheme), used in our analysis; see ref.~\cite{Blossier:2010cr}.}
\label{Qmassvaluse}
\end{table}

As can be seen in Table~\ref{simuldetails}, at $\beta = 3.90$ we have performed more extensive simulations, which enable us to check in some detail the quality and stability of the measured physical quantities. We wish to highlight straightaway the two problems we have encountered in these tests, performed for the tm-setup: (i) For all sea quark masses, when the valence quark attains its lightest value $a \mu_\ell =0.0040$, the vector meson effective mass has a poor plateau. The situation already improves at the next quark mass $a \mu_\ell =0.0064$. Nevertheless, since the signal-to-noise ratio behaves as expected (i.e. it drops like $\exp[-(m_{\rm V} - m_{\rm PS}) x^0]$) the $\rho$-meson mass and decay constant can still be extracted (see results presented in ref.~\cite{Jansen:2009hr}). (ii) A poor quality vector meson effective mass is also seen when $\mu_\ell < \mu_{\rm sea}$. This problem is absent in the pseudoscalar channel. 

\begin{figure}[!h]
\begin{center}
\subfigure[]{\includegraphics[scale=0.55]{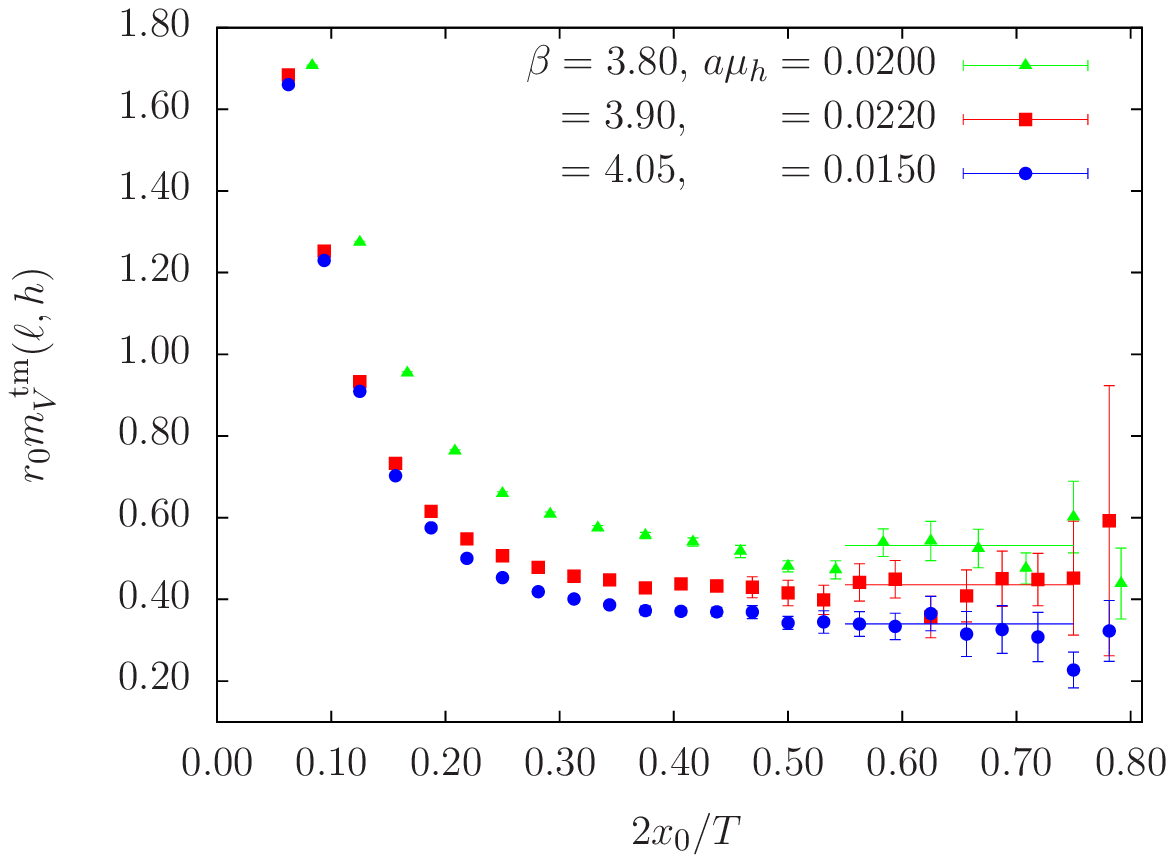}}
\subfigure[]{\includegraphics[scale=0.55]{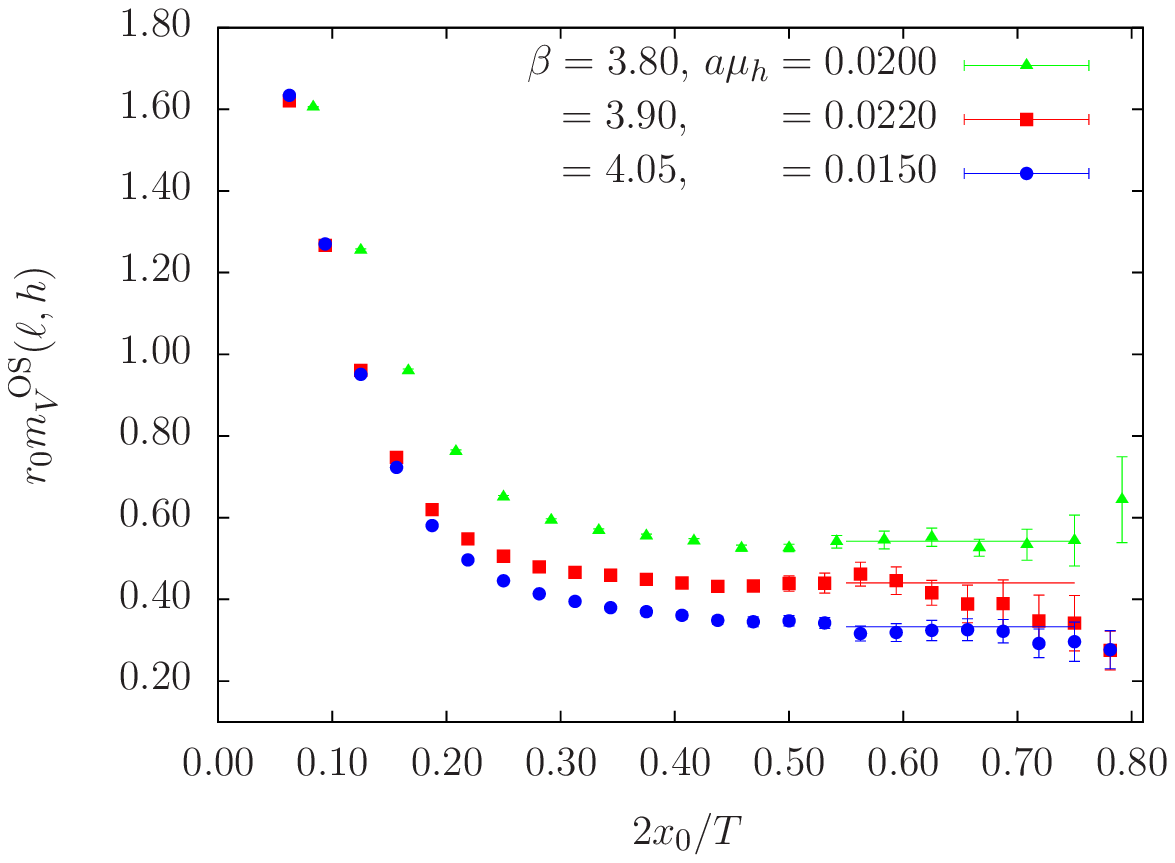}}
\caption{Effective vector meson mass $r_0 m_V$ at three values of the lattice spacing. The light quark mass is $a \mu_{\rm min}$ (see Table~\ref{simuldetails}) and the heavy quark mass $a \mu_h$ is close to that of the physical strange quark. (a) tm-setup ; (b) OS-setup. Plateau intervals are indicated by straight lines.}

\label{MV_plat}
\end{center}
\end{figure}

\begin{figure}[!h]
\begin{center}
\subfigure[]{\includegraphics[scale=0.55]{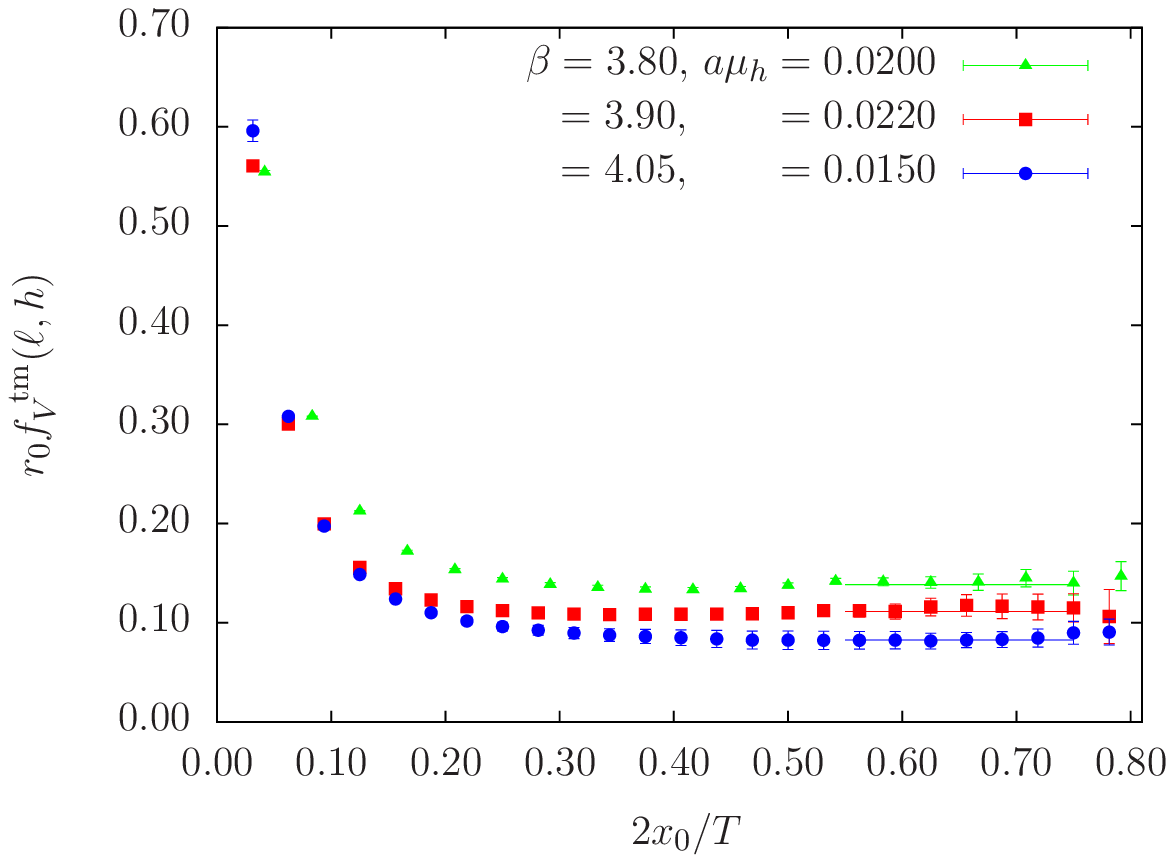}}
\subfigure[]{\includegraphics[scale=0.55]{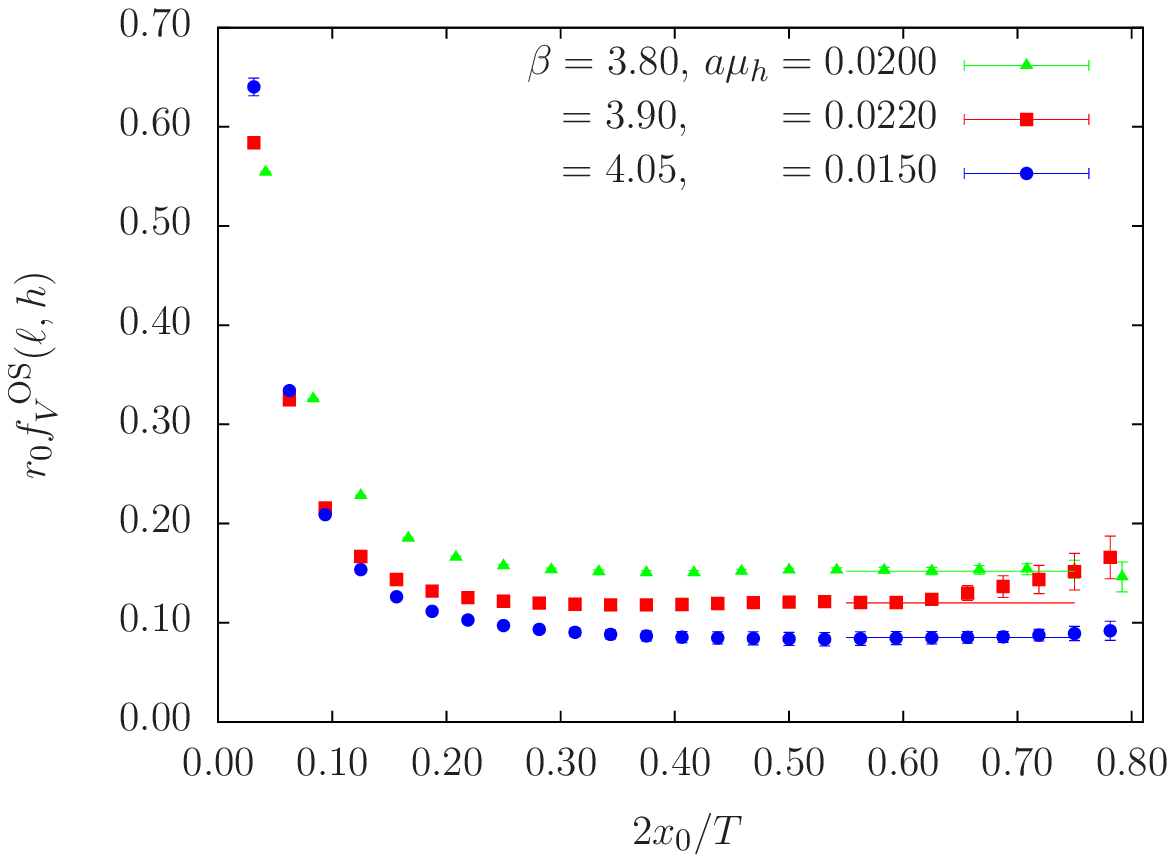}}
\caption{Vector decay constant $r_0 f_V$ at three values of the lattice spacing. The light quark mass is $a \mu_{\rm min}$ (see Table~\ref{simuldetails}) and the heavy quark mass $a \mu_h$ is close to that of the physical strange quark. (a) tm-setup ; (b) OS-setup. Plateaux intervals are indicated by straight lines.}
\label{FV_plat}
\end{center}
\end{figure}

The above problems are easily avoided in the present work, since the quantities of interest are related to the $\Kstar$-meson, consisting of a down and a strange valence quark mass ($\mu_{u/d} < \mu_s$). We thus proceed as follows: at each $\beta$ value, we compute the necessary observables (vector meson mass $m_V$, vector decay constant  $f_V$, and the ratio $f_T/f_V$), for all combinations of $a \mu_\ell = a \mu_{\rm sea}$ and $a \mu_h$ (with $\mu_\ell < \mu_h$). In this way unitarity holds in the light quark sector, while the heavy valence quark mass, in a partially quenched rationale, spans a range around the physical value $\mu_s$. Examples of the quality of our signal are given in Figs.~\ref{MV_plat} and \ref{FV_plat}; the lightest mass is $a \mu_{\rm min}$ and the heavy mass, corresponding to the physical strange value $a \mu_s$, is obtained by interpolation, as will be explained below.

\begin{figure}[!h]
\begin{center}
\subfigure[]{\includegraphics[scale=0.55]{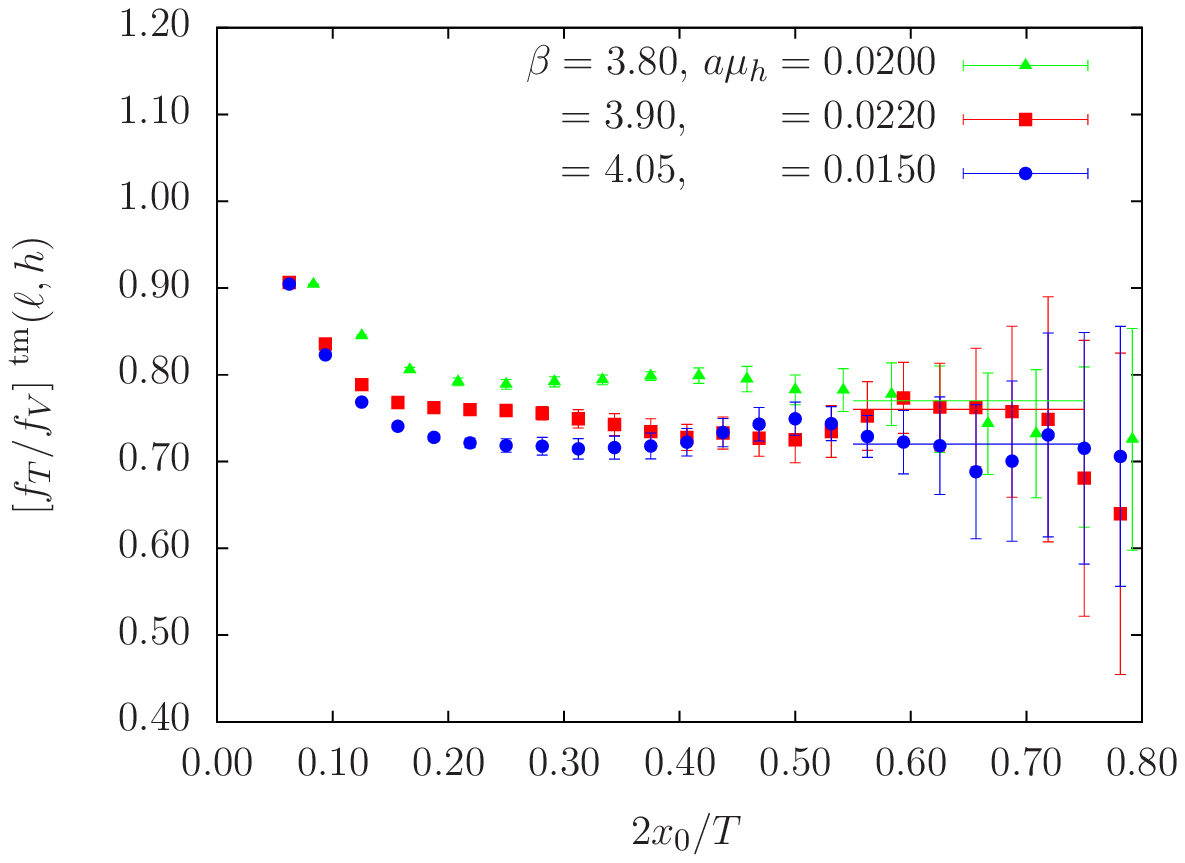}}
\subfigure[]{\includegraphics[scale=0.55]{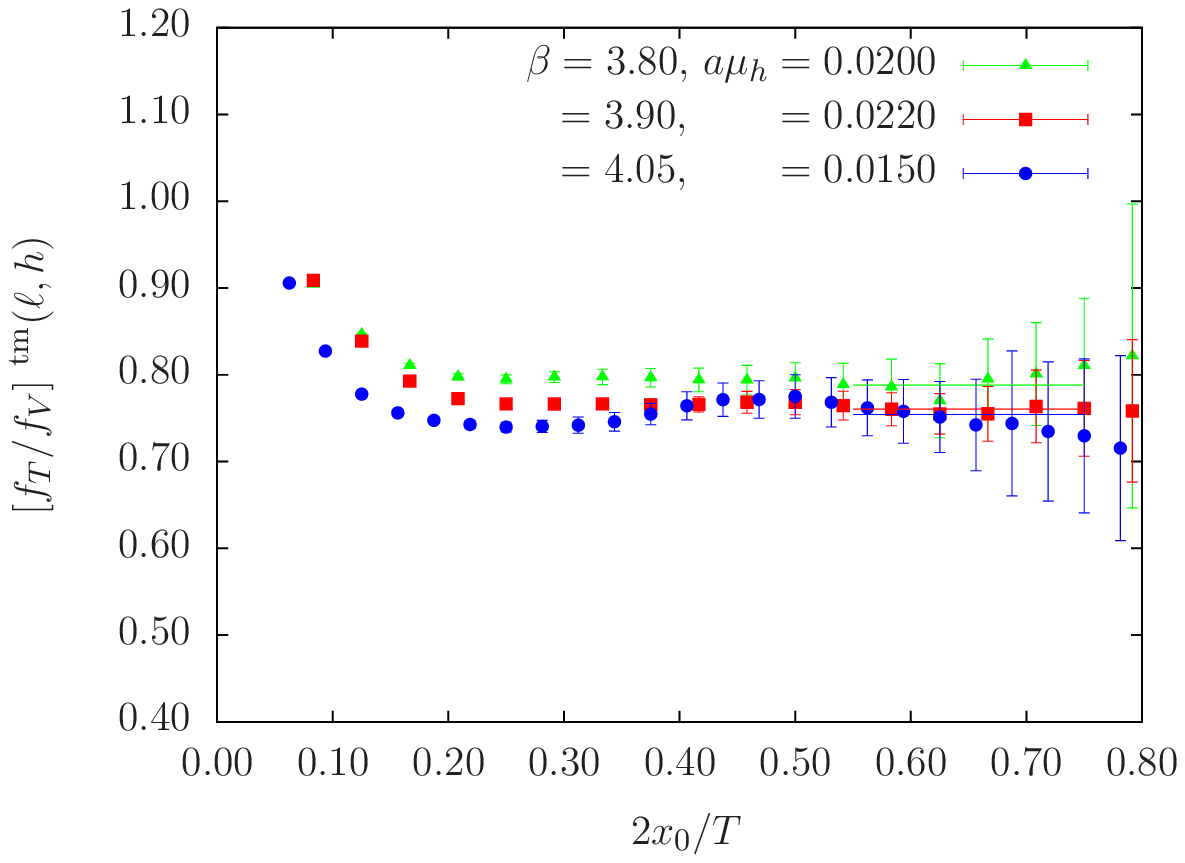}}
\caption{The ratio $f_T/f_V$ in the tm-setup, at three values of the lattice spacing and heavy quark mass $a \mu_h$, close to that of the physical strange quark. (a) For the lightest quark mass $a \mu_{\rm min}$; (b) for the next-to-lightest quark mass. Plateaux intervals are indicated by straight lines.}
\label{RAT_plat}
\end{center}
\end{figure}

Statistical errors are estimated with the bootstrap method, employing 1000 bootstrap samples. A reliable direct determination of the ratio $f_T/f_V$ in the OS-setup is not possible, because the ratio of correlation functions $C^{\rm OS}_{\tilde T}/C^{\rm OS}_V$ do not display  satisfactory plateaux, due to big statistical fluctuations of the tensor correlator $C^{\rm OS}_{\tilde T}$. We only present $f_T/f_V$ results in the tm-setup, obtained from the better-behaved correlation function $C^{\rm tm}_T$. In Fig.~\ref{RAT_plat} we show results for this ratio at $a \mu_{\rm min}$ and also at a heavier light quark mass.


Regarding vector meson masses $m_V$ and couplings $f_V$, both tm- and OS-results display similar plateau quality and statistical accuracy. At  finite lattice spacing and for equal bare quark masses, tm- and OS-estimates of  $m_V$ are compatible within errors. Agreement is also very good for $f_V$, with occasional discrepancies, interpreted as cutoff effects, showing up at the coarsest lattice\footnote{Given the large fluctuations of $f_T/f_V$ in the OS-setup at the finer lattice spacing, we only quote results for this ratio in the tm-setup.}.
Contrary to the well known large ${\cal O}(a^2)$ isospin breaking effects in the neutral to charged pion splitting mass, no numerically large differences are observed between tm and OS results for $f_V$ and $m_V$. This fact is in agreement with theoretical expectations, see ref.~\cite{Dimopoulos:2009qv}.

\begin{table}[!h]
\begin{center}
\scalebox{0.75}{
\begin{tabular}{cclccccc}
\hline \hline
&& & & & & & \\
$\beta$ && $a\mu_{l}$ & $r_0 m_V^{\rm{tm}}(\ell,s)$  &  $r_0 m_V^{\rm{OS}}(\ell,s)$ & 
$r_0f_{V}^{\rm{tm}}(\ell,s)$ & $r_0 f_{V}^{\rm{OS}}(\ell,s)$ & $[f_{T}/f_{V}]^{\rm{tm}}(\ell, s)$ \\
&& & & & & & \\
\hline
&&&&&&& \\
3.80 &&0.0080             &2.443(41)&2.471(30)&0.642(18)&0.700(13)&0.764(38)	   \\
&&0.0110                  &2.508(32)&2.500(23)&0.651(14)&0.706(15)&0.792(35)	   \\
&&&&&&& \\
\hline
&&&&&&& \\
3.90&&0.0040              &2.410(41)&2.381(38)&0.610(21)&0.643(17)&0.755(19)	\\
&&0.0064                  &2.441(32)&2.427(35)&0.626(22)&0.659(12)&0.726(20)	\\
&&0.0085                  &2.484(48)&2.441(33)&0.628(16)&0.652(16)&0.776(27)	\\
&&0.0100                  &2.468(54)&2.481(32)&0.619(20)&0.657(16)&0.774(31)	\\
&&0.0030(L=32)        
                          &2.259(75)&2.335(45)&0.577(20)&0.639(16)&0.714(20)   \\
&&0.0040(L=32)   
                          &2.364(32)&2.371(50)&0.599(22)&0.640(21)&0.722(19)   \\
\hline
&&&&&&& \\
4.05&&0.0030              &2.305(86)&2.263(80)&0.568(49)&0.588(40)&0.742(27)	  \\
&&0.0060                  &2.439(67)&2.295(76)&0.618(41)&0.578(46)&0.768(30)	  \\
&&0.0080                  &2.512(65)&2.427(48)&0.649(31)&0.648(27)&0.741(31)	  \\
\hline
&&&&&&& \\
CL && $\mu_{u/d}$     &2.227(71)&2.200(60)&0.545(41)&0.525(30)&0.701(46)  \\  
&&&&&&& \\
expt. &&                   &\hspace{1.6cm} 2.025&&\hspace{1.6cm} 0.493&&  \\          
\hline \hline
\end{tabular}
}
\end{center}
\caption{Results for three values of lattice spacing and several light quark masses  $a\mu_{\ell}$, interpolated to the physical strange mass $\mu_s$. Vector mass and vector decay constant results are presented for both  tm- and OS-setups . The ratio $f_{T}/f_{V}$ results are given only in the tm-setup. Our extrapolations at the  $\mu_{u/d}$ physical point and in the continuum limit are also shown. In the last row the experimental results for the vector mass and the vector decay constant, in units of $r_0$, have been added.  }
\label{results}
\end{table}

The extrapolation to the physical quark masses is carried out in two steps. First, for fixed values of the gauge coupling $\beta$ and  light quark mass $a \mu_\ell = a \mu_{\rm sea}$, we perform linear interpolations of $r_0 m_V$, $r_0 f_V$ and $f_T/f_V$ to the physical strange quark mass $\mu_s$. The second step consists in using these interpolated results for a combined fit of our data at three lattice spacings and all available light quark masses, in order to determine the continuum value of the quantity of interest ($r_0 m_V$, $r_0 f_V$ and $f_T/f_V$). The fitting function we use is
\be
m_V r_0 \,\, = \,\, C_0(\mu_s r_0 ) \,\, + \,\,  C_1(\mu_s r_0) \, \mu_\ell r_0  \,\, + \,\,  D(\mu_s r_0 ) \, \dfrac{a^2}{r_0^2} \,\,\, ,
\label{eq:comb-fit}
\ee
and similarly for $f_V r_0$ and $f_T/f_V$. The results of the interpolations in the heavy quark mass $\mu_h$ to the physical value $\mu_s$, at each $\beta$ and $a\mu_\ell$, are gathered in Table~\ref{results}. In the same Table we also display the results of the combined chiral and continuum extrapolations. Note that for the three quantities of interest, $m_V, f_V$ and $f_T/f_V$, the value of $\chi^2/{\rm d.o.f.}$ is less than unity. The linear dependence of our data on the light quark mass agrees with the predictions of  chiral perturbation theory for the ratio $f_T/f_V$ in the $\Kstar$ mass range; see~refs.\cite{Cata:2007ns,Cata:2009dq}.

Our final results, extracted in the tm-setup, are
\begin{eqnarray}
m_{\Kstar} \,\, &=& 981(31)(10)[33] {\rm MeV} \,\,\, , \\
f_{\Kstar} \,\, &=& 240(18)(02)[18] {\rm MeV} \,\,\, .
\end{eqnarray}
The first error includes the statistical uncertainty and the systematic effects related to the simultaneous chiral and continuum fits, mass interpolations and extrapolations, and uncertainties in the renormalization parameters. The second error arises from that of $r_0$. These two errors, combined in quadrature, give the total error in the square brackets. It is encouraging that these results agree with the ones obtained in the OS-setup (which is a different regularization), namely $m_{\Kstar} = 969(27)(10)[29] {\rm MeV}$ and $f_{\Kstar} = 231(13)(02)[13] {\rm MeV}$. Compared to the experimentally known values, $m_{\Kstar} = 892 {\rm MeV}$ and $f_{\Kstar} = 217 {\rm MeV}$, the vector meson mass is 2-3 standard deviations off, while the decay constant is compatible within about one standard deviation.

Our final estimate (tm-setup) for the ratio of vector meson couplings is
\be
\dfrac{\fT(2\, { \rm GeV})}{f_V}\Big \vert_{\Kstar} \,\, = \,\, 0.704(41) \,\,\, .
\ee
This is compatible with the continuum limit quenched result $[\fT(2 \,{\rm GeV})/f_V]_{\Kstar} = 0.739(17)(3)$ 
of ref.~\cite{Becirevic:2003pn}.
We are also in agreement with the result of the RBC/UKQCD collaboration~\cite{Allton:2008pn}; 
using $N_f = 2+1$ dynamical fermions at a single lattice spacing,  they quote $[f_T(2 \,{\rm GeV})/f_V]_{\Kstar} = 0.712(12)$. The lattice results are also in agreement with the sum rules' estimate $[f_T(2 \,{\rm GeV})/f_V]_{\Kstar} = 0.73(4)$, quoted in~\cite{Ball:2006eu}.

\begin{figure}[!h]
\begin{center}
\subfigure[]{\includegraphics[scale=0.55]{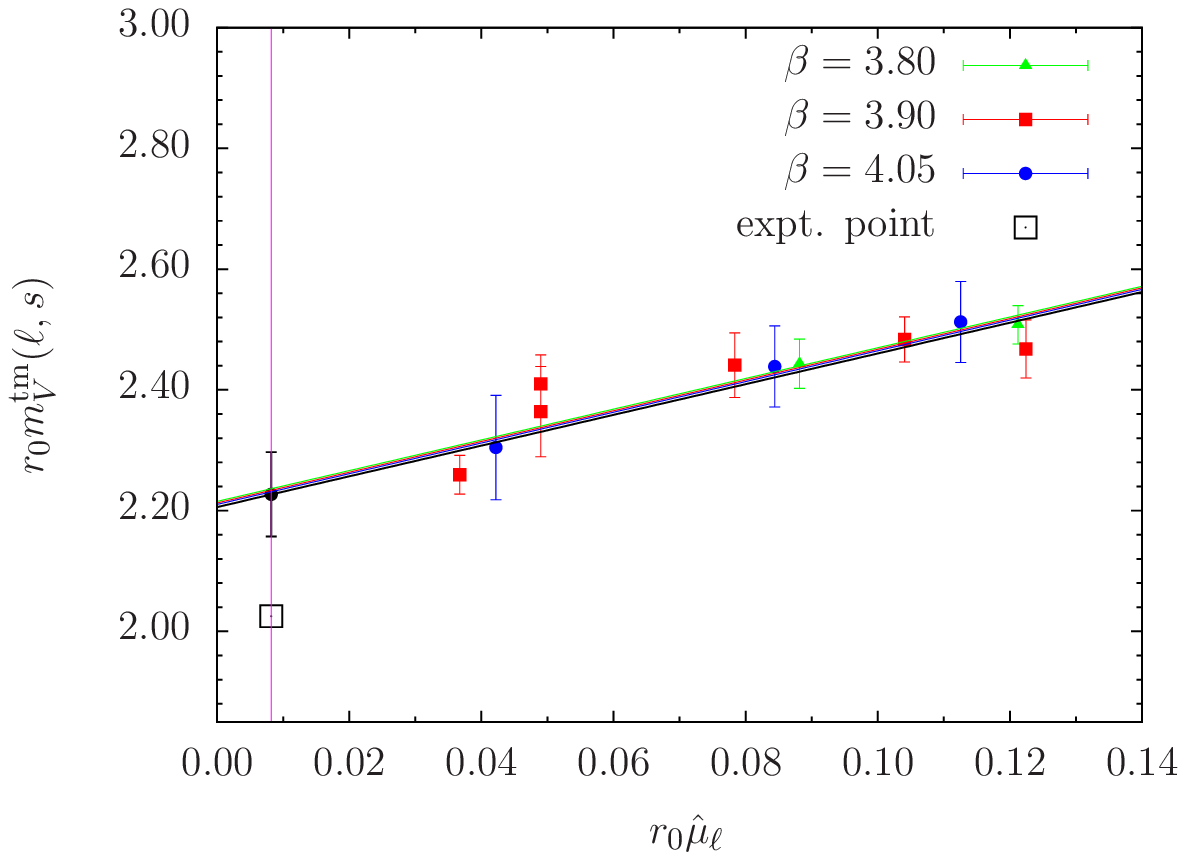}}
\subfigure[]{\includegraphics[scale=0.55]{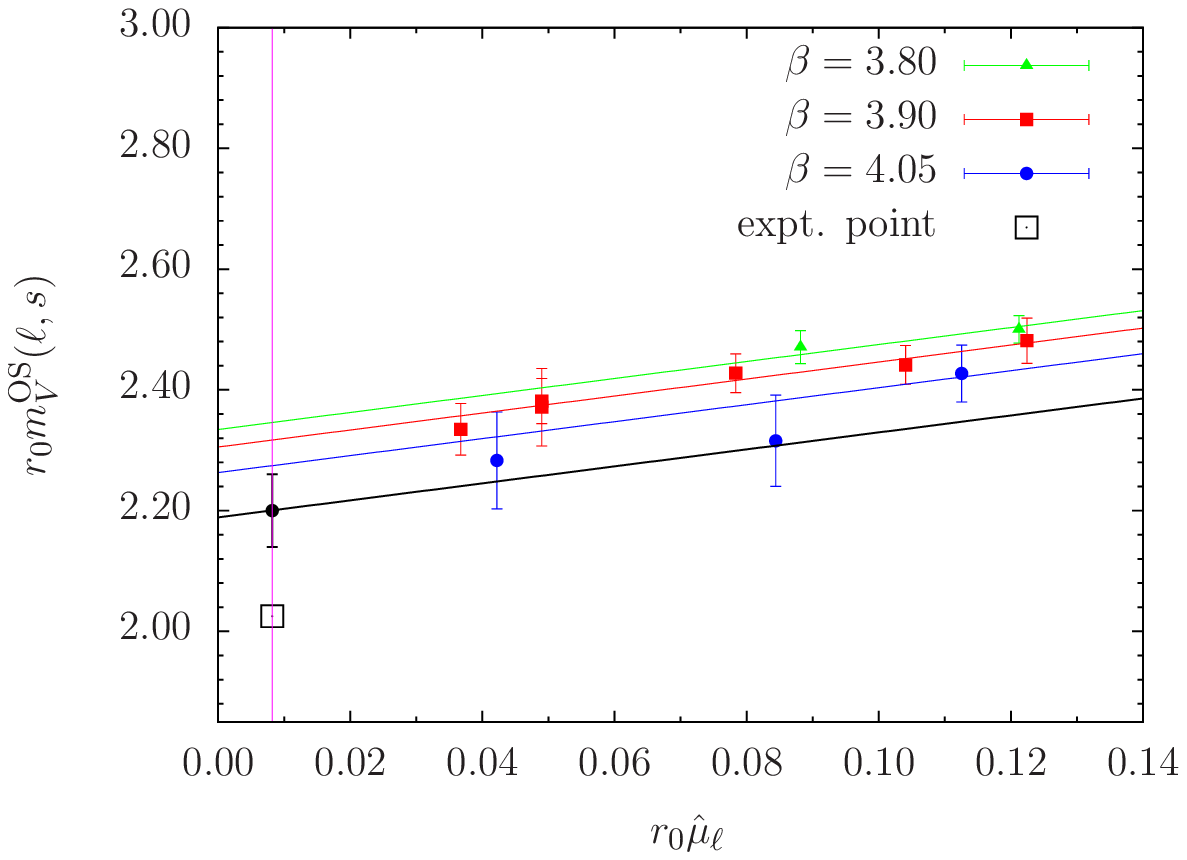}}
\caption{$r_0 m_V$ plotted against the renormalized light quark mass $r_0 \hat \mu_\ell$ ; (a) tm-setup; (b) OS-setup.
The continuous lines are combined chiral and continuum extrapolations to the physical point. 
The bottom (black) line corresponds to eq.~(\ref{eq:comb-fit}) at $a=0$. The  separation among the four lines in (a) is invisible to the naked eye (i.e. small scaling violations).}
\label{MV-cont}
\end{center}
\end{figure}

\begin{figure}[!h]
\begin{center}
\subfigure[]{\includegraphics[scale=0.55]{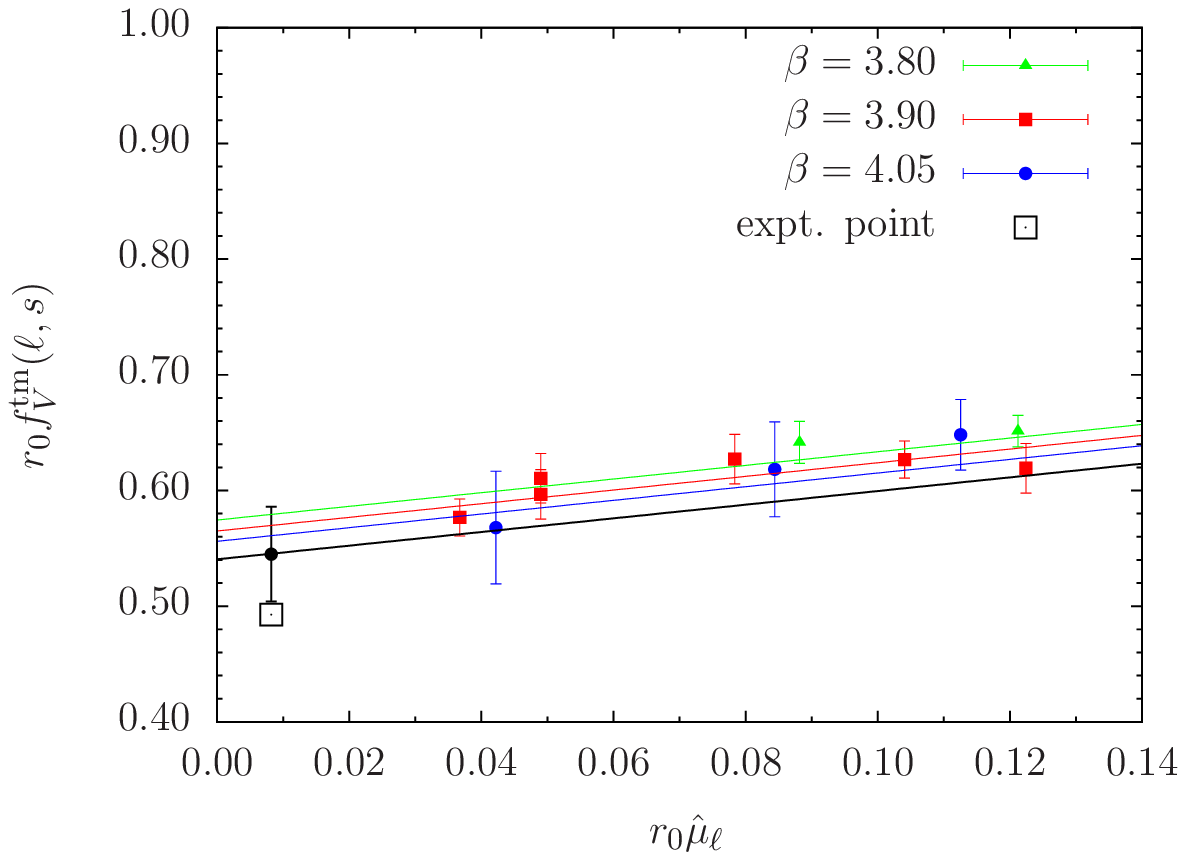}}
\subfigure[]{\includegraphics[scale=0.55]{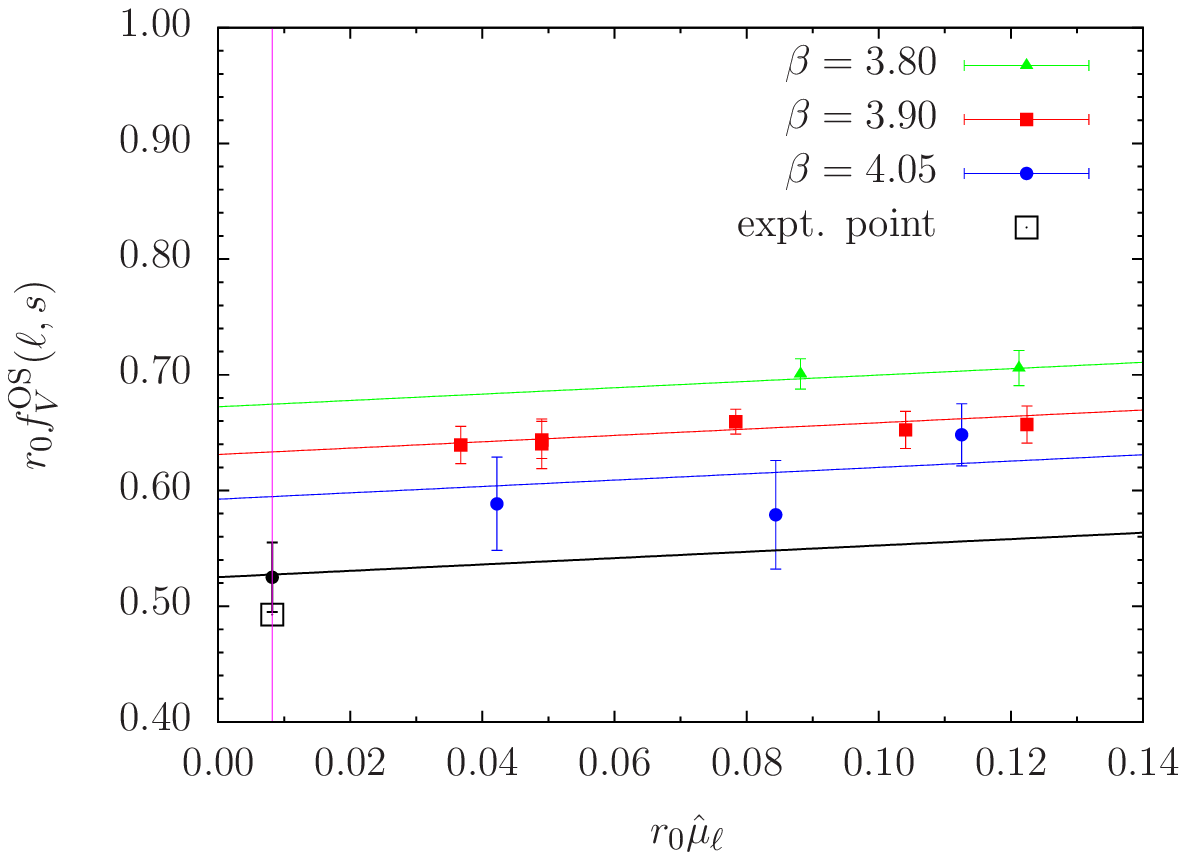}}
\caption{$r_0 f_V$ plotted against the renormalized light quark mass $r_0 \hat \mu_\ell$ ; (a) tm-setup; (b) OS-setup.
The continuous lines are combined chiral and continuum extrapolations to the physical point. 
The bottom (black) line corresponds to eq.~(\ref{eq:comb-fit}) at $a=0$.}
\label{FV-cont}
\end{center}
\end{figure}

\begin{figure}[t!]
\begin{center}
\includegraphics[scale=0.65]{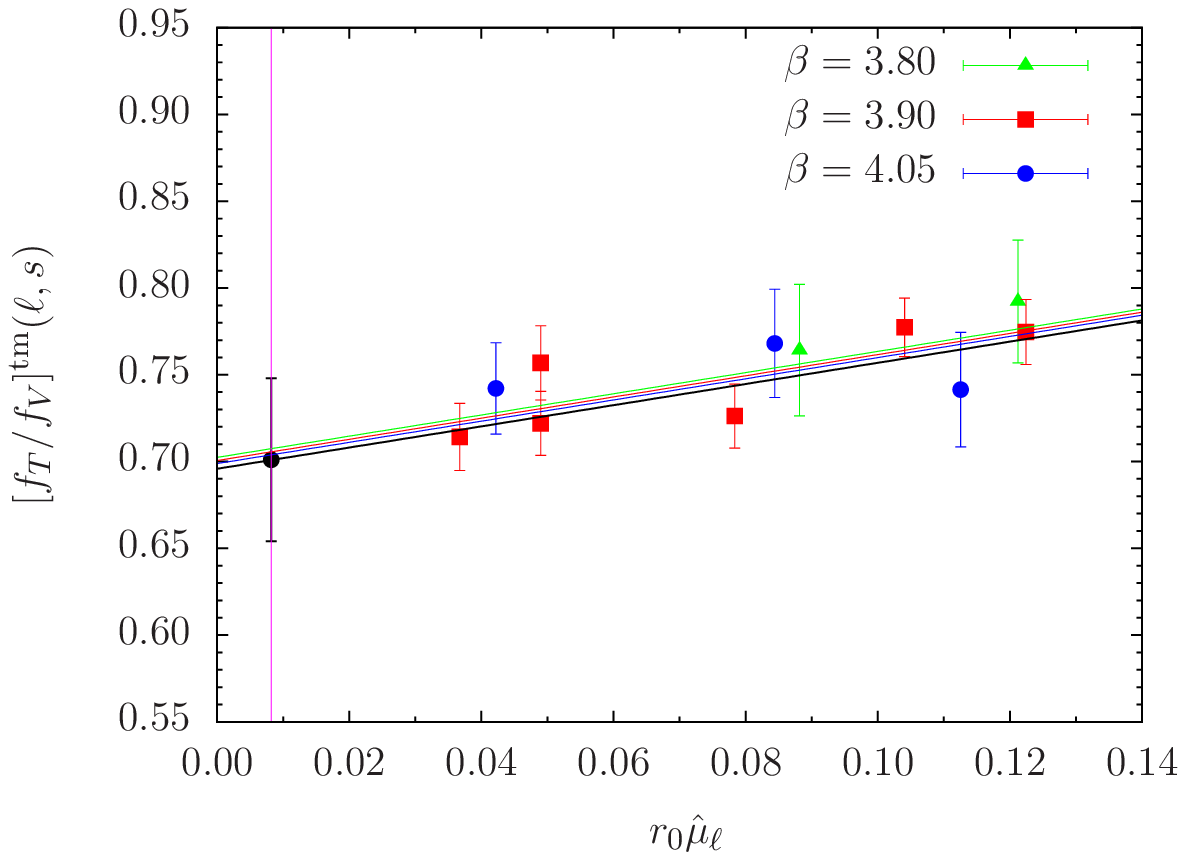}
\caption{$f_T/f_V$ plotted against the renormalized light quark mass $r_0 \hat \mu_\ell$ in the tm-setup.
The continuous lines are combined chiral and continuum extrapolation to the physical point. 
The bottom (black) line corresponds to eq.~(\ref{eq:comb-fit}) at $a=0$. The four lines are almost indistinguishable (i.e. small scaling violations).}
\label{RAT-cont}
\end{center}
\end{figure}

\section* {Acknowledgements}

{We thank G.C.~Rossi and C.~Tarantino for having carefully read the manuscript and for their useful comments and suggestions. We acknowledge  fruitful collaboration with  all ETMC members. We have greatly benefited from discussions with O.~Cata, C.~Michael, C.~McNeile, S.~Simula and N.~Tantalo.  F.M. acknowledges the financial support from projects FPA2007-66665, 2009SGR502, Consolider CPAN, and CSD2007-00042.
\pagebreak

\bibliography{lattice}        

\providecommand{\href}[2]{#2}\begingroup\begin{thebibliography}{10}

\bibitem{Baron:2009wt}
{\bf ETM} Collaboration, R.~Baron {\em et ~al.}, ``{Light Meson Physics from
  Maximally Twisted Mass Lattice QCD}'',  {\em JHEP} {\bf 1008} (2010) 097,
  [\href{http://xxx.lanl.gov/abs/0911.5061}{{\tt 0911.5061}}].

\bibitem{Frezzotti:2000nk}
{\bf ALPHA} Collaboration, R.~Frezzotti {\em et ~al.}, ``{Lattice QCD with a
  chirally twisted mass term}'',  {\em JHEP} {\bf 08} (2001) 058,
  [\href{http://xxx.lanl.gov/abs/hep-lat/0101001}{{\tt hep-lat/0101001}}].

\bibitem{Osterwalder:1977pc}
K.~Osterwalder and E.~Seiler, ``{Gauge Field Theories on the Lattice}'',  {\em
  Ann. Phys.} {\bf 110} (1978) 440.

\bibitem{tmqcd:DIrule}
{\bf ALPHA} Collaboration, C.~Pena, S.~Sint, and A.~Vladikas, ``Twisted mass
  QCD and lattice approaches to the Delta I = 1/2 rule'',  {\em JHEP} {\bf 09}
  (2004) 069, [\href{http://xxx.lanl.gov/abs/hep-lat/0405028}{{\tt
  hep-lat/0405028}}].

\bibitem{Frezzotti:2004wz}
R.~Frezzotti and G.~C. Rossi, ``{Chirally improving Wilson fermions. II:
  Four-quark operators}'',  {\em JHEP} {\bf 10} (2004) 070,
  [\href{http://xxx.lanl.gov/abs/hep-lat/0407002}{{\tt hep-lat/0407002}}].

\bibitem{FrezzoRoss1}
R.~Frezzotti and G.~C. Rossi, ``Chirally improving Wilson fermions. I: O(a)
  improvement'',  {\em JHEP} {\bf 08} (2004) 007,
  [\href{http://xxx.lanl.gov/abs/hep-lat/0306014}{{\tt hep-lat/0306014}}].

\bibitem{Dimopoulos:2007cn}
{\bf ALPHA} Collaboration, P.~Dimopoulos {\em et ~al.}, ``{Flavour symmetry
  restoration and kaon weak matrix elements in quenched twisted mass QCD}'',
  {\em Nucl. Phys.} {\bf B776} (2007) 258--285,
  [\href{http://xxx.lanl.gov/abs/hep-lat/0702017}{{\tt hep-lat/0702017}}].

\bibitem{Dimopoulos:2009es}
{\bf ALPHA} Collaboration, P.~Dimopoulos, H.~Simma, and A.~Vladikas, ``{Quenched $B_K$-parameter from
  Osterwalder-Seiler tmQCD quarks and mass-splitting discretization effects}'',
   {\em JHEP} {\bf 07} (2009) 007,
  [\href{http://xxx.lanl.gov/abs/0902.1074}{{\tt 0902.1074}}].

\bibitem{Blossier:2010cr}
{\bf ETM} Collaboration, B.~Blossier {\em et ~al.}, ``{Average up/down, strange
  and charm quark masses with Nf=2 twisted mass lattice QCD}'',  {\em Phys.
  Rev.} {\bf D82} (2010) 114513, [\href{http://xxx.lanl.gov/abs/1010.3659}{{\tt
  1010.3659}}].

\bibitem{Constantinou:2010gr}
{\bf ETM} Collaboration, M.~Constantinou {\em et ~al.}, ``{Non-perturbative
  renormalization of quark bilinear operators with Nf=2 (tmQCD) Wilson fermions
  and the tree- level improved gauge action}'',  {\em JHEP} {\bf 08} (2010)
  068, [\href{http://xxx.lanl.gov/abs/1004.1115}{{\tt 1004.1115}}].

\bibitem{Jansen:2009hr}
{\bf ETM} Collaboration, K.~Jansen {\em et ~al.}, ``{Meson masses and decay
  constants from unquenched lattice QCD}'',  {\em Phys. Rev.} {\bf D80} (2009)
  054510, [\href{http://xxx.lanl.gov/abs/0906.4720}{{\tt 0906.4720}}].

\bibitem{Dimopoulos:2009qv}
{\bf ETM} Collaboration, P.~Dimopoulos, R.~Frezzotti, C.~Michael, G.~Rossi, and C.~Urbach, ``{O(a**2)
  cutoff effects in lattice Wilson fermion simulations}'',  {\em Phys.Rev.}
  {\bf D81} (2010) 034509, [\href{http://xxx.lanl.gov/abs/0908.0451}{{\tt
  0908.0451}}].

\bibitem{Cata:2007ns}
O.~Cata and V.~Mateu, ``{Chiral perturbation theory with tensor sources}'',
  {\em JHEP} {\bf 0709} (2007) 078,
  [\href{http://xxx.lanl.gov/abs/0705.2948}{{\tt 0705.2948}}].

\bibitem{Cata:2009dq}
O.~Cata and V.~Mateu, ``{Chiral corrections to the f(V)-perpendicular /f(V)
  ratio for vector mesons}'',  {\em Nucl.Phys.} {\bf B831} (2010) 204--216,
  [\href{http://xxx.lanl.gov/abs/0907.5422}{{\tt 0907.5422}}].

\bibitem{Becirevic:2003pn}
D.~Becirevic {\em et ~al.}, ``{Coupling of the light vector meson to the vector
  and to the tensor current}'',  {\em JHEP} {\bf 05} (2003) 007,
  [\href{http://xxx.lanl.gov/abs/hep-lat/0301020}{{\tt hep-lat/0301020}}].

\bibitem{Allton:2008pn}
{\bf RBC-UKQCD} Collaboration, C.~Allton {\em et ~al.}, ``{Physical Results
  from 2+1 Flavor Domain Wall QCD and SU(2) Chiral Perturbation Theory}'',
  {\em Phys.Rev.} {\bf D78} (2008) 114509,
  [\href{http://xxx.lanl.gov/abs/0804.0473}{{\tt 0804.0473}}].

\bibitem{Ball:2006eu}
P.~Ball, G.~W. Jones, and R.~Zwicky, ``{B $\rightarrow$ V gamma beyond QCD
  factorisation}'',  {\em Phys.Rev.} {\bf D75} (2007) 054004,
  [\href{http://xxx.lanl.gov/abs/hep-ph/0612081}{{\tt hep-ph/0612081}}].

\end{thebibliography}\endgroup

\end{document}